\documentclass[aps,floatfix,showpacs,amsmath,superscriptaddress,nofootinbib,
12pt,a4paper]{revtex4-1}
\usepackage{graphicx}

\begin{document}

\preprint{}

\title{Single scale factor for the universe from the 
creation of radiation and matter till the present}

\author{Recai Erdem}
\email{recaierdem@iyte.edu.tr}
\affiliation{Department of Physics,
{\.{I}}zmir Institute of Technology \\ 
G{\"{u}}lbah{\c{c}}e K{\"{o}}y{\"{u}}, Urla, {\.{I}}zmir 35430, 
Turkey} 

\date{\today}

\begin{abstract}
A scheme for incorporating the creation of radiation and 
matter into the cosmological evolution is introduced so that it becomes 
possible to merge the times before and after the creation of radiation and 
matter in a single scale factor in Robertson-Walker metric. This 
scheme is illustrated through a toy model that has the prospect of 
constituting a basis for a realistic model. 
\end{abstract}

\maketitle
\raggedbottom
\section{Introduction}

The question of determining the model that best describes the universe is 
the ultimate goal of cosmology. The energy-momentum content of the present 
universe seems to be a perfect fluid mainly consisting of a dark sector 
(possibly consisting of a dark energy and a dark matter component), 
baryonic matter, and radiation \cite{cosmology}. In the standard model of cosmology 
(namely, $\Lambda$CDM)  dark matter \cite{DM} and baryonic matter are 
considered to 
be dust, dark energy \cite{DE} is taken to be the Einstein's 
cosmological 
constant, 
and radiation is described by the usual energy-momentum term for 
radiation. Although the standard model seems to be 
compatible with 
observations yet it has some problems. The magnitudes of  potential 
theoretical contributions to cosmological constant (CC) 
are 
extremely higher than the value of CC deduced from 
the energy density of the universe \cite{CCP}. 
There are many 
attempts to solve this problem, namely, the CC problem. Nevertheless none 
is wholly satisfactory. 
The best option seems to employ a symmetry such as metric reversal 
symmetry \cite{MRS} to cancel CC and then attribute the dark energy to 
something 
else e.g. to modified gravity \cite{modified-grav}, or to some scalar 
field such as 
quintessence \cite{quintessence}. Cold dark matter (i.e. dust-like dark 
matter with 
no or negligible interaction with itself and with baryonic matter and 
photons) scenario of  
$\Lambda$CDM as well suffers from some problems such as rotation curves of 
spiral-like galaxies i.e. cuspy halo problem, missing satellite galaxies 
problem \cite{cuspy,satellite}. There are many alternatives to cold dark 
matter (CDM) scenario 
including warm-dark matter \cite{warm-DM}, Bose-Einstein condensate dark 
matter \cite{condense-DM,scalar-condensate}, and scalar field dark matter 
\cite{scalar-DM}. 

The above considerations essentially hold for the time from the 
radiation dominated era till the present era. The standard paradigm for 
the era before the radiation dominated era is an inflationary era (that 
serves to solve the problems of the standard cosmology 
such as horizon, flatness, absence of monopoles problems) 
\cite{inflation}. 
Usually the inflationary era and the epoch after this era are studied 
separately. This is not only due to the need to concentrate on each of 
these and to try to understand each epoch better before a 
possible unification. In fact the most serious problem in the direction of 
the unification 
\footnote{I mean a true unification i.e. description of 
the whole cosmological evolution by a single scale factor in 
the metric.}
of the whole cosmic history is the difficulty of merging 
these two epochs because of the form of the dependence of the energy 
density of dust and radiation on scale factor (i.e. on redshift). 
In $\Lambda$CDM the energy density of 
radiation dominates over that of inflaton 
if one goes back to sufficiently large redshifts.
This is due to the fact that the energy density of inflaton is essentially 
constant during 
inflationary era while the energy density of radiation scales like 
$\frac{1}{a^4}$ where $a$ is the scale factor. In other words, to have a 
true unification, the creation of radiation and 
matter after the inflationary era must be taken into account in the 
scale factor without destroying the standard cosmology before and after 
the inflation, and this is not an easy task. The models in 
literature that unify all eras of cosmological evolution in a single model 
\cite{all-unified,Erdem}
are not wholly realistic since they do not include baryonic matter 
although they are able to produce eras of cosmological 
evolution with correct equations of state in the corresponding eras, and 
some have graceful exit from 
inflationary era. The matter in these 
models must be identified with dark matter since the 
energy densities of these models do not contain energy components that 
scale 
proportional to $\frac{1}{a^3}$ for all times (or 
at least for a sufficiently long time). 
The models in 
\cite{all-unified} use the energy densities expressed in 
terms of simple functions of Hubble parameter 
and/or scale parameter as the starting point rather than starting
from the scale factor. 
Although one may, in principle, determine the scale factor from this 
information the form of scale factor may be rather complicated in 
some cases. On the 
other hand a relatively 
simple scale factor may result in a rather complicated and unmanageable 
functional form for the energy density when expressed in 
terms of the scale factor or the Hubble parameter. Therefore in some cases 
it 
may be more suitable to consider a specific ansatz for scale factor 
such as in this study and in \cite{Erdem}. The same approach is adopted in 
this study.
Moreover the present study introduces
a general prescription to include dust and radiation into unification.

In this study, first, in Section II, I introduce a scheme to unify the 
cosmological evolution
before and after the radiation dominated era. Then I give a 
concrete realization of this scheme in Section III. In Section IV I 
discuss the 
observational compatibility of this scheme in the context of the model 
introduced in Section III. Finally I conclude in Section V.   The scale 
factor in this model is a sum of two terms. The 
first term is a pure dark 
energy contribution. The second term is responsible for the 
baryonic matter and radiation terms and additional terms that may be 
mainly identified with dark matter. There is also an additional term due 
to coupling 
between these terms, and this term gives another contribution to 
the dark energy and dark matter. 
Some of the ideas employed here have been already studied 
in literature. 
In this study I do not make a sharp distinction between dark energy and 
dark matter because the dark energy and dark matter terms are coupled 
and the equation of state (EoS) of some terms 
e.g. EoS of the coupling term between dark matter and dark energy terms 
evolve with time. 
The superficiality of 
a distinction between dark energy and dark matter is considered in 
many studies in literature, either explicitly or implicitly 
\cite{Matos1, Matos2, Arbey,unified,Aviles}. 
This option is quite possible 
since dark energy and dark matter are not observed directly. What we see 
observationally is a missing element in the energy-momentum tensor of the 
Einstein equations, other than baryonic matter and radiation, and this 
missing quantity may be described by two components; dark energy and dark 
matter. It is, in principle, equally possible that this quantity is 
composed of a single component, say, dark fluid. 
In \cite{Erdem} I had introduced a universe composed of a
dark fluid (that may be written in terms of two scalar fields). In fact 
the scale factor in that study is essentially $a_1(t)$ in Eq.(\ref{a2}) of 
this paper. 
The present study, somewhat, may be considered 
as an extension of \cite{Erdem} where baryonic matter and radiation are 
included. However there are important differences as well. The main aim 
of this 
study is to introduce a scheme to merge the cosmological evolution of the 
time before and after the production of radiation into a single scale 
factor with the baryonic matter and the usual radiation terms included. 
The modified form of $a_1(t)$ in \cite{Erdem} only serves 
as a realization of 
this scheme. Furthermore I do not discuss the 
scalar field identification of the energy density due the part of 
the scale factor similar to $a_1(t)$ of \cite{Erdem} 
(although it can be easily done), and I do not consider the 
cosmological perturbations of 
these quantities, and the inflationary era in this study because these 
points would cause divergence of the main goal of the paper
and would increase the volume of this study drastically. I leave 
these points to future studies.

\section{Outline of the Model}

Consider the Robertson-Walker metric
\begin{equation}
ds^2\;=\;g_{\mu\nu}dx^\mu\,dx^\nu\,=\,-dt^2\,+\,a^2(t)\tilde{g}_{ij}dx^idx^j
\label{a1} 
\end{equation}
I take the 3-dimensional space be flat, i.e. 
$\tilde{g}_{ij}=\delta_{ij}$ for the sake of simplicity, which is an 
assumption consistent with cosmological observations \cite{PDG,Planck}. 
I let the form of the scale factor be
\begin{eqnarray}
a(t)&=&c_0\left(a_1(t)\,-\,a_2(t)\right) \label{a2} \\
&&
c_0=\frac{1}{A_1-A_2}~,~~A_1=a_1(t_0)~,~~A_2=a_2(t_0) \nonumber
\end{eqnarray}
where $t_0$ denotes the present time. We will see that $a_1(t)$ is the 
part of 
the scale factor responsible for 
dark energy and dark matter, and $a_2(t)$ is the one 
mainly responsible for dust and radiation and additional contribution to 
dark matter-energy, and we shall see later that a 
mixing between the sectors due to $a_1$ and $a_2$ act as an additional 
source of dark energy. We assume that $a_1(t)$ and $a_2(t)$ are chosen in 
such a way that $a(t)\,>\,0$ for all $t$. In general one may  
identify the dust by a mixture of baryonic matter
and dust-like dark matter. The best fit values that I could find 
by trial and error for the specific toy model considered in this 
study for 
implementation of the present scheme 
seem to prefer  
the case where the dust 
term is wholly or almost wholly  due to baryonic matter. 

We first focus on the 
$a_2(t)$ term and specify it as 
\begin{equation} 
a_2(t)\,=\,x(t)\,a(t) \label{a3} 
\end{equation}
where $x(t)$ is some function that its form will be specified later.
Eqs. (\ref{a2}) and (\ref{a3}) may be used to relate 
$a(t)$ and $a_1(t)$, $a_2(t)$ in a more applicable way, and to derive the 
corresponding Hubble parameter.
We observe that
\begin{eqnarray}
a_2&=&
ax\,=\,c_0(a_1-a_2)x~~~\Rightarrow~~
a_2=\frac{c_0x}{1+c_0x}a_1~,~~
a\,=\,\frac{c_0}{1+c_0x}a_1\,=\,\frac{1}{x}a_2
\label{a4a}
\end{eqnarray}
In a similar way the Hubble parameter is found to be
\begin{eqnarray}
&&\frac{\dot{a}_2}{a}\,=\,\frac{\dot{a}}{a}\,+\,\dot{x}~~~\Rightarrow~~~
H\,=\,\frac{\dot{a}}{a}\,=\,
\frac{\dot{a}_{1n}}{a_{1n}}
-\frac{c_0\dot{x}}{1+c_0x}
\label{a4b}
\end{eqnarray}
where we have used
\begin{eqnarray}
\frac{\dot{a}_1}{a}
&=&\left(\frac{1+c_0x}{c_0}\right)\frac{\dot{a}_1}{a_1}~,~~
\frac{\dot{a}_{1}}{a_{1}}\,=\,
\frac{\dot{a}_{1n}}{a_{1n}}
\label{a5a}\\
&&a_1\,=\,A_1\,a_{1n}~,~~a_{1n}(t_0)=1 \nonumber 
\label{a5b}
\end{eqnarray}
Note that $a(t_0)=1$
by convention.

We let 
\begin{eqnarray} 
&&x(t)\,=\,A_2\frac{1}{c_1-c_2}x_3(t)\left(c_1x_1(t)-c_2x_2(t)\right) 
\label{a3a} \\ 
&&x_1(t)\,=\,\exp{\int_{t_0}^t\tilde{H}_2^{(1)}dt}~,~~x_2(t)\,=\, 
\exp{\int_{t_0}^t\tilde{H}_2^{(2)}dt} \label{a3b} \\ 
&&x_3(t)\,=\, \exp{\int_{t_0}^t\tilde{H}_2^{(3)}dt} \label{a3c}
\end{eqnarray} 
where $c_1$, $c_2$ are some constant coefficients, and 
\begin{eqnarray} 
&&\tilde{H}_2^{(1)} 
\,=\,\alpha_{o1}\left[-\frac{\alpha_b}{a^{\frac{3}{2}}}- 
\frac{\alpha_r}{a^2}+\frac{\alpha_x}{a^3}\right]\,,~ 
\tilde{H}_2^{(2)} 
\,=\,\alpha_{o2}\left[\frac{\alpha_b}{a^{\frac{3}{2}}}+ 
\frac{\alpha_r}{a^2}-\frac{\alpha_K}{a}\right]\,,~ 
\tilde{H}_2^{(3)} 
\,=\,\alpha_c\left(\frac{1}{a^3}-\frac{1}{a}\right) \nonumber \\
\label{a10a} \end{eqnarray} 
where $\alpha_{o1}$, $\alpha_{o2}$, $\alpha_b$, $\alpha_r$, $\alpha_x$, 
$\alpha_K$ are some other constant coefficients.
In fact, 
in (\ref{a3a}) we could take the simpler form where $\alpha_c=0$, 
$\alpha_{o1}=\alpha_{o2}=1$, $c_1=1$, 
$c_2=0$. This 
would be enough as long as we are concerned only with merging of the eras 
before and after the radiation domination, and the resulting model would 
be compatible with Union2 data set at an order of magnitude level. The 
more involved form in (\ref{a3a}) is used to make the model 
phenomenologically more viable. 
This point will 
be discussed when we discuss the phenomenological viability of the model 
in Section IV. One may determine $\dot{x}$ in Eq.(\ref{a4b}) by using Eq.(\ref{a3a}), 
\begin{equation} 
\dot{x}(t) 
\,=\,A_2\frac{1}{c_1-c_2}x_3\left[-B(t)\left(\frac{\alpha_b}{a^\frac{3}{2}}+ 
\frac{\alpha_r}{a^2}\right)\,+\,S(t)\frac{\alpha_x}{a^3}\,+\,K(t)
\frac{\alpha_K}{a}\right]
\end{equation}
where
\begin{eqnarray}
B(t)&=&\frac{A_2}{c_1-c_2}x_3
\left(c_1\alpha_{o1}x_1+c_2\alpha_{o2}x_2\right) \label{c1} \\
S(t)&=&\frac{A_2}{c_1-c_2}x_3\left[c_1\alpha_{o1}\alpha_xx_1+
\alpha_c\left(c_1x_1-c_2x_2\right)\right]
\label{c2} \\
K(t)&=&\frac{A_2}{c_1-c_2}x_3\left[c_2\alpha_{o2}\alpha_Kx_2-
\alpha_c\left(c_1x_1-c_2x_2\right)\right] \label{c3}
\end{eqnarray}
Hence
one may express 
(\ref{a4b}) as
\begin{equation}
H\,=\,H_{1n}\,+\,A(t)\,\tilde{H}_2\,+\,H_\Delta
\label{a3ac}
\end{equation}
where
\begin{eqnarray}
H_\Delta&=&-\Xi(t)\,\frac{1}{a^3}\,-\,\psi(t)\,\frac{1}{a} \label{a3da}\\
&&A(t)\,=\,
\frac{c_0B(t)}{1+c_0x(t)}~,~~
\tilde{H}_2\,=\,
\frac{\alpha_b}{a^{\frac{3}{2}}}\,+\, 
\frac{\alpha_r}{a^2} \label{a3db} \\
&&\Xi(t)\,=\,
\frac{c_0S(t)}{1+c_0x(t)}~,~~
\psi(t)\,=\,
\frac{c_0K(t)}{1+c_0x(t)} \label{a3dc}
\end{eqnarray}

We let
\begin{equation}
H_{1n0}=\tilde{\Omega}_1^\frac{1}{2}H_0~,~~
A_0\alpha_b=\tilde{\Omega}_b^{\frac{1}{2}}H_0~,~~
\,A_0\alpha_r=\tilde{\Omega}_r^{\frac{1}{2}}H_0~,~~
\,\Xi_0=\tilde{\Omega}_x^{\frac{1}{2}}H_0~,~~
\,\psi_0=\tilde{\Omega}_K^{\frac{1}{2}}H_0
\label{a10b}
\end{equation}
where $H_{1n0}=H_{1n}(t_0)$, $H_0=H(t_0)$,
$A_0=A(t_0)$, $\Xi_0=\Xi(t_0)$, $\psi_0=\psi(t_0)$. Because the three dimensional part 
of metric is taken to be flat the present energy density is equal to the 
critical energy density the above equations imply that
\begin{equation}
\tilde{\Omega}_1^\frac{1}{2}\,+\,
\tilde{\Omega}_b^{\frac{1}{2}}\,+\,
\tilde{\Omega}_r^{\frac{1}{2}}\,-\,
\tilde{\Omega}_x^{\frac{1}{2}}\,-\,
\tilde{\Omega}_K^{\frac{1}{2}}\,=\,1
\label{a10ba}
\end{equation}
Note that, at this point
$\tilde{\Omega}_1$,
$\tilde{\Omega}_b$,
$\tilde{\Omega}_r$,
$\tilde{\Omega}_x$,
$\tilde{\Omega}_K$ cannot be identified as density parameters since
density parameters should satisfy
$\Omega_1$+$\Omega_b$+$\Omega_r$+$\Omega_x$+$\Omega_K$=1. In Chapter IV we 
will see that this condition is not satisfied for the phenomenologically viable 
sets of parameters, so
$\tilde{\Omega}_1$,
$\tilde{\Omega}_x$,
$\tilde{\Omega}_K$ cannot be identified as density parameters separately, 
instead one must define the total density parameter for dark sector by 
$\Omega_D^\frac{1}{2}$=$\tilde{\Omega}_1-
\tilde{\Omega}_x-\tilde{\Omega}_K$ rather than the separate contribution 
due to $H_{1n}$ and $H_\Delta$ while I identify 
$\tilde{\Omega}_b$, $\tilde{\Omega}_r$ as the density 
parameters corresponding dust and radiation.
Therefore to retain the 
physical content of this paper more evident I will not make a distinction 
between 
$\tilde{\Omega}_b$, $\tilde{\Omega}_r$ 
and the density parameters 
for baryonic matter, radiation; $\Omega_b$, $\Omega_r$  
while I keep this distinction for the others i.e. for the ones due 
to $H_{1n}$ and $H_\Delta$ terms.
The $\frac{\alpha_b}{a^\frac{3}{2}}$ and
$\frac{\alpha_r}{a^2}$ terms result in energy densities that are 
identified as the energy densities for baryonic matter and radiation. In 
principle, there may be also  contributions due to the 
$\Xi\frac{1}{a^3}$ and $\psi\frac{1}{a}$. 
The sign of the $\Xi\frac{1}{a^3}$ term is negative of 
the usual stiff matter. It may be identified as stiff matter under 
pressure so that it has a negative deceleration parameter. The main 
function of 
this term is to damp the 
energy densities of baryonic matter 
and radiation in the time before the radiation dominated era. The function 
of the $\frac{1}{a}$ term is similar. It ensures the behavior of the 
energy density in late times be well-behaved (i.e. preventing the energy 
density to grow too 
fast (through the $\frac{1}{a}$ term in $x_2(t)$ and $x_3(t)$)). 
Although the $\psi\frac{1}{a}$ term is similar to that of a negative 
curvature 3-space it is different from such a term since its origin is 
the Hubble parameter $H$ while a usual 3-curvature term arises from the 
3-curvature part of metric. Note that this term arises even in a flat 
3-space 
in this construction.  Therefore I identify the $\frac{\Xi}{a^3}$ and 
$\frac{\psi}{a}$ terms in $H$ as additional contributions to dark sector.

Another point worth to mention is; It is evident that the 
square of (\ref{a4b}) (in conjunction with
(\ref{a10a})) results in an $A^2\tilde{H}_{2}^2$ 
term containing  $A^2\frac{\alpha_b^2}{a^3}$ and
$A^2\frac{\alpha_r^2}{a^4}$ terms that may be identified with 
the standard baryonic matter and radiation terms, respectively 
if $A$ is taken to be constant while it depends on time in this scheme as 
it is evident from (\ref{a3a}). In fact variation of $A$ with time makes 
it possible to go to zero before the radiation dominated era as desired. 
Therefore, given the considerable success of the standard model at least 
in the observed relatively small redshifts, the variation in $A$ after the 
matter - radiation decoupling time should be small so that this scheme 
mimics the standard model at relatively small redshifts where 
observational data is available. If one takes
$\left(\frac{dA}{dt}\right)_{t\simeq\,t_0}$ sufficiently small one may 
guarantee an almost constant value for $A$ for a sufficiently long 
time (e.g. from the present time till the beginning of the radiation 
dominated era). We will see in Section IV 
that there exist such values of $A$ with reasonable phenomenological 
viability.
Another term 
arising from $\tilde{H}_2^2$ is the cross term, 
$A^2\frac{\alpha_b\alpha_r}{a^\frac{7}{2}}$. This term may be identified 
as the energy density term due to the transitory time where massive 
particles that act as radiation at high energies turn into more dust-like 
at intermediate energies. Another term in $H^2$ is $H_{1n}^2$. This term 
will be considered as a pure dark sector term. Finally the cross term 
$2H_{1n}\tilde{H}_2$ gives an additional contribution to the dark sector 
for the phenomenologically viable values of the parameters. 
It may be easily shown that 
this term does not necessarily imply strong interaction between the 
dark fluid and 
radiation and baryonic matter as its form may suggest 
if the parameters of the underlying physics at microscopic 
scale satisfy some restrictions. Otherwise one may use 
screening mechanisms such as 
\cite{chameleon,symmetron,Pospelov} to explain the 
unobservablity of dark matter-energy.

Next we derive the general form of 
the equation of state for this 
model. We derive the explicit form of the equation of state after (EOS) 
after we give the explicit form of $a_1(t)$ in the section. 
However giving the general form of EOS in this scheme provides us a more 
model independent formula and may be useful for other choices of $a_1(t)$ 
in future. 
After 
using Eqs.(\ref{a3ac},\ref{a3da},\ref{a3db},\ref{a3dc}) 
one obtains EOS, $\omega$ as
\begin{eqnarray}
\omega&=&\frac{p}{\rho}\,=\,\frac{\frac{G_{11}}{g_{11}}}{G_{00}}\,=\,
-\frac{2\dot{H}+3H^2}{3H^2} \nonumber \\
&=&
-\frac{2\dot{H}_{1n}\,+\,3H_{1n}^2}{3H^2}
\,-\,\frac{AH_{1n}\left(\frac{3\alpha_b}{a^\frac{3}{2}}+\frac{2\alpha_r}{a^2}
\right)}{3H^2}\,+\,\frac{A^2\left(\frac{\alpha_r^2}{a^4}
+\frac{\alpha_b\alpha_r}{a^\frac{7}{2}}\right)}{3H^2} \nonumber \\
&&\,-\,\frac{2\dot{A}\tilde{H}_2}{3H^2}
\,+\,
\frac{
A\left(\frac{\Xi}{a^3}+\frac{\psi}{a}\right)
\left(\frac{3\alpha_b}{a^\frac{3}{2}}+\frac{2\alpha_r}{a^2}\right)}{3H^2}
\,+\,
\frac{6H_{1n}\left(\frac{\Xi}{a^3}+\frac{\psi}{a}\right)}{3H^2}\nonumber 
\\
&&-\,
\frac{2H\left(\frac{3\Xi}{a^3}+\frac{\psi}{a}\right)\,+\,3
\left(\frac{\Xi^2}{a^6}+\frac{\psi^2}{a^2}+\frac{2\Xi\psi}{a^4}\right)}
{3H^2}
\label{a11a}
\end{eqnarray}
The terms inside the first parenthesis in the second line correspond to 
the contribution of 
the dark sector term $H_{1n}$. The other
terms in the same line correspond to the 
contributions of dust and radiation and their coupling with 
dark sector term $H_{1n}$. The remaining terms 
are the term corresponding to variation of $A$, 
the term corresponding to coupling of 
curvature-like term and the stiff matter under 
negative pressure with dust and radiation, the 
term corresponding to coupling of 
curvature-like term and the stiff matter under negative 
pressure with $H_{1n}$, the term corresponding to coupling of 
curvature-like term and the stiff matter under negative pressure 
with the other terms, and the contribution of the curvature-like 
term and the stiff matter under negative pressure, respectively.
It is evident from (\ref{a11a})
that the pressure for baryonic matter is zero as should be, and the 
pressure for radiation is $\frac{1}{3}$ as 
expected.
A point worth to mention at this point 
is; The coupling term between baryonic matter and radiation 
in Eq.(\ref{a11a}) has an equation of state $\frac{1}{6}$ (that may be 
seen by considering the ratio of the 
$\frac{\alpha_b\alpha_r}{a^\frac{7}{2}}$ in $p$ by the corresponding term 
in $\rho$ i.e. $2\frac{\alpha_b\alpha_r}{a^\frac{7}{2}}$). The redshift 
dependence of this term is between that of baryonic matter and 
radiation. This time dependence is more natural than the standard picture 
where there is no such term. 
Massive particles at high energies act as radiation 
and at lower turns into dust. The coupling term accounts for the 
transitory time when massive particles pass from radiation to dust state. 

\begin{figure}
\includegraphics[scale=0.9]{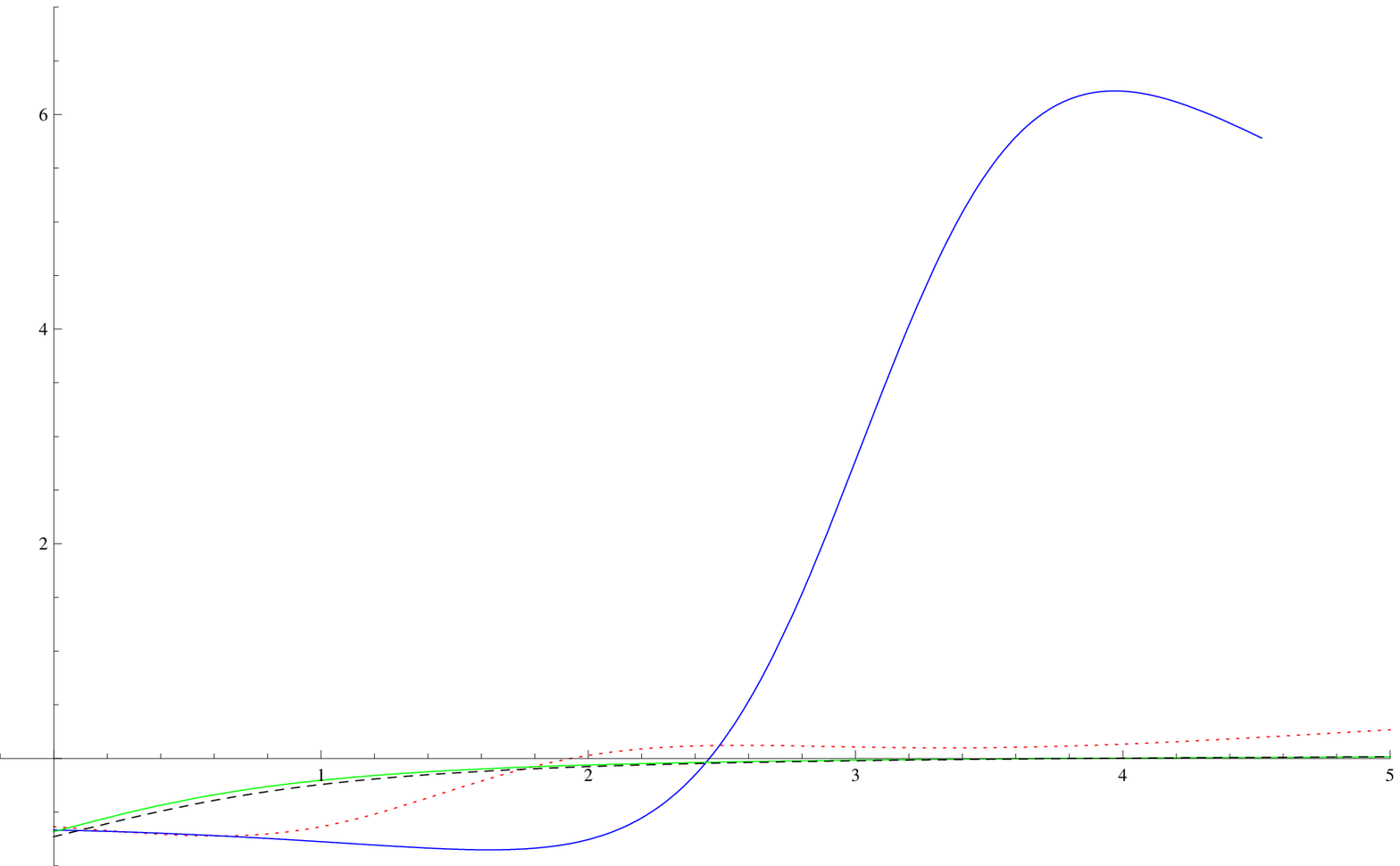}
\caption{
$\omega$ versus redshift z graphs for $\Lambda$CDM (for PDG and Planck values) and for this model 
(for two sets of parameters with small $\chi^2$ values), namely,
for $\Lambda$CDM with 
$\Omega_\Lambda=0.6825$, 
$\Omega_b=0.3136$,  
$\Omega_r=1.3\times\,10^{-3}$  
(solid green) and with 
$\Omega_\Lambda=0.73$, 
$\Omega_b=0.2661$, 
$\Omega_r=1.3\times\,10^{-3}$  
(dashed black); and for this model
for the set of parameters 
in Table 9 (solid blue) and for the set of parameters 
($r$=1.58, $s$=5.3, $\beta$=3.1, $\Omega_b^\frac{1}{2}$=0.21,
$\xi\i_1$=0.975, $A_1=c_1=\alpha_{o2}$=1, $A_2$=$10^{-4}$, 
$c_2$=0.999, $\frac{\alpha_r}{\alpha_b}$=0.1, 
$\alpha_{ac}$=0.05, $\alpha_{o1}$=0.9, 
$\frac{\alpha_x}{\alpha_b}$=
$\frac{\alpha_K}{\alpha_b}$=0.7) (dotted red).
Here the sub-index $b$ refers to dust.}
\label{fig:1}
\end{figure}

In order to obtain the evolution of $\omega$ as a 
function  redshift or time explicitly, 
$H_{1n}$ must be specified. This will be done in the next section.
However I give a $\omega$ versus redshift graph in Fig.\ref{fig:1}
for $a_{1n}$ introduced in the next section for a phenomenologically
viable set of parameters (i.e. those with small $\chi^2$ values and 
with energy densities for recombination and nucleosynthesis as 
discussed in Section IV) to have an idea about the evolution of
$\omega$ with redshift. To draw this graph I have converted time, 
t to redshift, z (for Union2.1 data) through 
the relation $z=\frac{1}{a}-1$, and then used Mathematica to use 
this relation to make the calculations (although the original 
quantities are expressed in terms of time). This procedure is
applicable for small redshifts. However, in general,
it becomes inapplicable 
due to highly non-linear form of scale factor and Hubble parameter 
since it requires huge RAM and CPU for computation, if 
it can be done at all,  
and hence requires a separate computational physics project by 
itself. Therefore I have used equation of state versus and energy 
density versus time graphs (instead of redshift) in Section IV. 
In fact, even that option 
required a long time of order of months to make the 
necessary computations. 

\section{An Explicit Realization of the Model}

Now we focus on the $a_1(t)$ term. We take
\begin{eqnarray}
a_1(t)&=&A_1a_{1n}(t)\label{a13a} \\
a_{1n}(t)&=&[p_1\,+\,p_2b_2t]^r\,\exp{[-b_1(b_2t)^{-1/s}]} \label{a13b} 
\end{eqnarray}
where $A_1\,<\,1$,
$p_1$, $p_2$, $b_2$, $b_1$ are some constants that to be fixed or 
bounded by consistency arguments or cosmological observations. This scale 
factor is a generalization of the scale factor in 
\cite{Erdem} where 
$r=1$, $s=6$. A similar scale factor is considered in \cite{Sami} as 
well. One of the shortcomings of \cite{Erdem}
is that the present value of the equation of state parameter in that model 
(for phenomenologically relevant choices of parameters where the model mimics 
$\Lambda$CDM)
is $\sim\,-0.4$ while observations imply that it 
should be $\simeq\,(-0.68)$  ---  $(-0.74)$ \cite{PDG,Planck}. 
In the present study there 
is an 
additional contribution due to mixing of the terms due to $a_1$ and $a_2$ 
and hence there is less need to modify the scale factor in \cite{Erdem}. 
However
I prefer to adopt the more general form in (\ref{a13b}) to seek a greater 
parameter space and to insure the correct equation of state parameter.

We had shown in Eq.(\ref{a4b}) that the Hubble parameter may be expressed 
as 
$H\,=\,H_{1n}\,+\,A(t)\,\tilde{H}_2\,+\,H_\Delta$. 
Now we concentrate on the $H_{1n}=\frac{\dot{a}_{1n}}{a_{1n}}$ part of the 
Hubble parameter. In fact this amounts to specifying the model wholly 
since the other terms, as well, depend on $a_{1n}$ as we have seen.
The corresponding $H_{1n}$ is given by
\begin{equation}
H_{1n}(t)\,=\,
\frac{\dot{a}_{1n}}{a_{1n}}
\,=\,\frac{rp_2b_2}{(p_1+p_2b_2t)}
+\frac{1}{s}b_2b_1(b_2t)^{-\left(1+\frac{1}{s}\right)}
\label{a14} 
\end{equation}
We let
\begin{eqnarray}
&&1=a_{1n0}=a_{1n}(t_0)\,=\,
(p_1+p_2b_2t_0)^r
\exp{[-b_1(b_2t_0)^{-\frac{1}{s}}]}\label{a15} 
\end{eqnarray}
and 
\begin{eqnarray}
&&
H_{0}t_0\,=\,\xi,~~~
(p_1+p_2b_2t_0)^r \,=\,
\exp{[b_1(b_2t_0)^{-\frac{1}{s}}]}\,=\, \beta\,>\,1
\label{a16a} \\
H_{1n0}t_0
&&=\,H_{1n}(t_0)t_0\,=\,\frac{rp_2b_2t_0}{(p_1+p_2b_2t_0)}
+\frac{1}{s}b_1(b_2t_0)^{-\frac{1}{s}} 
\,=\,\xi\xi_1\,\nonumber \\
&&\Rightarrow~~~~
rp_2b_2t_0\,=\,\beta^{\frac{1}{r}}(\xi\xi_1-\frac{1}{s}\ln{\beta})~,~~~
p_1\,=\,
\beta^{\frac{1}{r}}[1-\frac{1}{r}(\xi\xi_1-\frac{1}{6}\ln{\beta})] 
\label{a16b} 
\\
&&\Rightarrow~~~
H_1(t)\,=\,H_{1n}(t)\,=\,H_{1n}(\gamma)
\,=\,\frac{1}{t_0}\{\frac{\xi\xi_1-\frac{1}{s}\ln{\beta}}
{[1+\frac{\gamma-1}{r}(\xi\xi_1-\frac{1}{s}\ln{\beta})]}
+\frac{1}{s}\gamma^{-\frac{s+1}{s}}\ln{\beta}\}  
\label{a16c} \\
&&\mbox{here}~~~~\gamma=\frac{t}{t_0} \nonumber
\end{eqnarray}
where $t_0$ is the present age of the universe. One observes from 
(\ref{a10b}) and the above expression
that
\begin{equation}
H_{1n0}=\tilde{\Omega}_1^\frac{1}{2}H_0=\frac{1}{t_0}\xi\xi_1 
~~\Rightarrow~~\tilde{\Omega}_1^\frac{1}{2}\,=\,\xi_1
\end{equation}
We will see in the next section that 
$\tilde{\Omega}_1$ 
cannot be identified as the density parameter 
corresponding to $H_{1n}$. Instead one must define an overall density
parameter for the dark sector by 
$\Omega_D^\frac{1}{2}$=
$\tilde{\Omega}_1^\frac{1}{3}-
\tilde{\Omega}_x^\frac{1}{3}-
\tilde{\Omega}_K^\frac{1}{3}$.  
Observational values 
of $H_0=\left(\frac{\dot{a}}{a}\right)_{t=t_0}$ and $\frac{1}{t_0}$ are 
almost the same. Therefore $\xi^2\xi_1^2\,\simeq\,\xi_1^2$.

After determining the $H_{1n}$ we are almost ready to find the explicit 
values of the energy density and the equation of state. The only missing 
element for calculation of these quantities is to find $A$, $\Xi$, $\psi$ 
in (\ref{a3ac}, \ref{a3da}, \ref{a3db}, \ref{a3dc}). Another point to 
be addressed is to show that there exist sets of A whose variation with 
time are small for low redshifts so that the terms that are proportional 
to $\frac{1}{a^\frac{3}{2}}$    
and $\frac{1}{a^2}$ in $A\,\tilde{H}_2$ term may be identified with dust and 
usual radiation terms, respectively.

In order to determine $A$, $\Xi$, $\psi$ (and to determine the rate of 
variation of $A$ with time) one should derive an approximation scheme 
for the evaluation of these quantities because these quantities depend on
$x_1(t)$,
$x_2(t)$,
$x_3(t)$ (that are defined in (\ref{a3b}) and (\ref{a3c})), and these 
quantities, in turn, are defined in a recursive way since 
$x_i(t)$=$\exp{(\int_{t_0}^t\tilde{H}_2^{(i)}dt)}$ ($i=1,2,3$) and 
$\tilde{H}_2^{(i)}$ depend on $a(t)$, and $a(t)$, in turn, depends 
on $x_i(t)$ through Eq.(\ref{a4a}). In other words, in order to determine 
the approximate values of $x_i(t)$ one must identify the zeroth order 
approximation and a method how to obtain the higher order approximations 
in an iterative way. One may use the following observations to obtain the
zeroth order approximation; 
$A_1c_0(1+c_0x(0))^{-1}=A_1\frac{1}{A_1-A2}(1+\frac{A_2}{A_1-A_2})^{-1}=1$ 
and $\dot{A}\sim\,0$ $\Leftrightarrow$ 
$\dot{x}_i\sim\,0$ ($i=1,2,3$) i.e. $\dot{x}\sim\,0$, $x(t)\simeq\,x(0)=1$ for 
small redshifts. This implies that the zeroth order approximation for the scale
factor $a(t)$ should be taken as $a^{(0)}(t)\,=\,a_{1n}(t)$
Hence for phenomenologically 
viable cases (where $\dot{A}\sim\,0$ for small redshifts) one may take 
the zeroth order approximations as 
\begin{equation} 
x_i^{(0)}(t)\,=\,\exp{\int_{t_0}^t\tilde{H}_2^{(i0)}dt}~~~i=1,2,3
\label{a3ba}
\end{equation} 
where 
$\tilde{H}_2^{(10)}$, $\tilde{H}_2^{(20)}$, $\tilde{H}_2^{(30)}$ is 
obtained from $\tilde{H}_2^{(1)}$, $\tilde{H}_2^{(2)}$, 
$\tilde{H}_2^{(3)}$ by replacing $a(t)$ by $a_{1n}(t)$ in those 
expression, for example, 
\begin{equation} \tilde{H}^{(10)}_2 
\,=\,\alpha_{o1}\alpha_b\int_{t_0}^t\{-\frac{1}{a_{1n}^{\frac{3}{2}}}- 
\frac{\frac{\alpha_r}{\alpha_b}}{a_{1n}^2}+ 
\frac{\frac{\alpha_x}{\alpha_b}}{a_{1n}^3}\} \label{a17a} \end{equation} 
Then 
\begin{eqnarray} 
&&A^{(0)}(t)\,=\, 
\frac{c_0B^{(0)}(t)}{1+c_0x^{(0)}(t)} \label{d1} \\ &&x^{(0)} 
(t)\,=\,\frac{A_2}{c_1-c_2}x_3^{(0)} 
\left(c_1x_1^{(0)}-c_2x_2^{(0)}\right) \label{d2} ~~etc.
\label{d2} \end{eqnarray} 
One may get the next order approximation by using 
\begin{equation}
a(t)\simeq\,a^{(1)}(t)= c_0A_1(1+c_0x^{(0)})^{-1} a_{1n}
\end{equation}
The next order quantities  $A^{(1)}$, $x^{(1)}$ may be obtained from (\ref{d1}) and (\ref{d2}) by replacing the 
superindices $(0)$ by $(1)$
where 
\begin{equation} 
x_i^{(1)}(t)\,=\,\exp{\int_{t_0}^t\tilde{H}_2^{(i1)}dt}
\label{a3bc} 
\end{equation}
Here $\tilde{H}_2^{(i1)}$ is obtained from 
$\tilde{H}_2^{(i)}$ by replacing $a(t)$ by 
$a^{(1)}(t)= c_0A_1(1+c_0x^{(0)})^{-1} a_{1n}$.
For k'th approximation we replace $a(t)$ by 
$a^{(k)}(t)= c_0A_1(1+c_0x^{(k-1)})^{-1} a_{1n}$.
In principle this may be done up to arbitrarily higher order approximations 
but it is quite difficult to calculate even $A^{(1)}$ even with the help 
of computers. In fact I have divided the interval $t\,-\,t_0$ in to 
coarser subintervals to decrease the CPU time and have used the 
approximate numerical values in the i'th interval (by assuming $A^{(0)}$ 
to be almost constant in those intervals) by using the formula 
\begin{equation} A^{(0)}(t_i)=\frac{A^{(0)}(t_{i-1})\,+\, 
A^{(0)}(t_{i+1})}{2} \label{ax}
\end{equation} 
to find $A^{(1)}$. I have seen (by trial and error) that it is 
possible to find almost constant
$A^{(0)}$ and $A^{(1)}$ values for many relevant (i.e. of 
small $\chi^2$ values considered in the next section) 
choices of parameters,
$\alpha_b$, $r$, $s$, $\xi_1$, $\xi_1$, $A_1$, $A_2$, $c_1$, $c_2$, 
$\alpha_r$, $\alpha_c$, $\alpha_{o1}$, $\alpha_{o2}$, $\alpha_x$, 
$\alpha_K$. For example the variations of  
$A^{(0)}$ and $A^{(1)}$ with time for one of the phenemonologically
viable sets in Table \ref{table-9} is given in Table \ref{table-1}.

\section{Compatibility with Observations}

Now we check the phenomenological viability of the model. The 
observational analysis of the model for all possible values of the 
parameters, $\beta$, $r$, $s$, $\xi$, $\xi_1$, etc. is an extremely 
difficult job (if not impossible at all) because expressing the Hubble 
parameter, deceleration parameter etc. in terms of the scale factor is 
quite difficult since these quantities are highly nonlinear 
functions of the scale factor in this model. Therefore I adopt 
some guidelines to seek the phenomenologically viable sets of 
parameters. These guidelines are: \\
1- I take the model mimic the standard model i.e. the $\Lambda$CDM 
model, at least from the time of decoupling of matter and radiation up to 
the present time. Therefore I take the present time values of 
the equation of state of the whole universe and the density parameter of 
the baryonic matter and radiation to be the same as  
$\Lambda$CDM. \\
2- In searching for the phenomenologically viable parameter space I start 
from the values of the parameters in \cite{Erdem} i.e. $r=1$, $s=6$, 
$\xi=1$, and $\beta\sim\,O(1)$ since the universe studied in \cite{Erdem} 
mimics the true universe roughly. \\
3- Due to the highly non-linear relation between the Hubble parameter and 
the scale factor I seek the relevant parameter space usually by trial and 
error rather than a continuous scan of the parameter space. Therefore the 
optimum values obtained here most probably may not correspond to the best 
possible optimization. Rather they hopefully correspond to a good 
approximation to the best optimal values.   

\subsection{Compatibility with Union2.1 Data}

In this subsection we use the Union2.1 compilation data set to 
find the optimal values of $\beta$, $r$, $s$ starting from 
$\beta=3$, $r=1$, $s=6$. We find the theoretical 
values of distance moduli, $\mu$ for the redshift values of 
Union2.1 and calculate the corresponding $\chi^2$ value 
by using the measured values of $\mu$ and their errors. 

The expression for distance modulus is 
\begin{equation}
\mu\,=\,5\,Log_{10}\left(\frac{d_L}{1\,Mpc}\right)\,+\,25 
\end{equation}
where
\begin{equation}
d_L\,=\,\frac{c\,a_0}{a(t)}\int_t^{t_0}\frac{dt^\prime}{a(t^\prime)}\,=\,
\frac{c\,a_0}{\frac{A_1c_0}{1+c_0x(t)}a_{1n}(t)}
\int_t^{t_0}\frac{dt}{\frac{A_1c_0}{(1+c_0x(t^\prime))}a_{1n}(t^\prime)} 
\label{mu1a}
\end{equation}
where for small redshifts reduces to
\begin{equation}
d_L\,
 \,\simeq\,
\frac{c}{a_{1n}(t)}\int_t^{t_0}\frac{dt^\prime}{a_{1n}(t^\prime)} 
\label{mu1b}
\end{equation}
where we have used the requirement that 
$\frac{A_1c_0}{1+c_0x}\simeq\,1$ at small redshifts as 
discussed in the preceding section (see Table \ref{table-7}), and 
$a_0=a(0)=1$. In $\Lambda$CDM 
$\int\frac{dt}{a(t)}$ is usually expressed in terms of redshift, $z$ and 
Hubble 
parameter $H$, and then the 
results for different $z$'s are compared with the data directly. This 
is not possible in this model because $H$ cannot be expressed in terms of  
$a(t)$ in a simple way. 
Therefore in this study first we convert redshift values of Union2 to time 
values by using $z=
\frac{1}{a(\gamma)}-1\simeq\, 
\frac{1}{a_{1n}(\gamma)}-1$ 
and then solve it for $\gamma$. The corresponding expression for the 
theoretical value of the luminosity distance $d_L$ in this case (i.e. in 
terms of $\gamma$) is
\begin{equation}
d_L\,\simeq\,
\frac{c\,t_0\,
\beta^{-1+\gamma^{-\frac{1}{s}}}}{[1+\frac{\gamma-1}{r}\left(\xi\xi_1
-\frac{1}{s}\ln{\beta}\right)]^r}
\int_\gamma^1 
d\gamma\frac{\beta^{-1+\gamma^{-\frac{1}{s}}}}
{[1+\frac{\gamma-1}{r}\left(\xi\xi_1-\frac{1}{s}\ln{\beta}\right)]^r}
\label{mu2}
\end{equation}
where $a_{1n}(t)$ is expressed in terms 
of $\beta$, $r$, $s$, $\gamma=\frac{t}{t_0}$ by using the parameterization 
given in the preceding section. Eq.(\ref{mu2}) may be written in more 
standard form 
in terms of $H_0$ by using $H_0t_0=\xi$. 
Then we find (\ref{mu1b}) numerically for 
each of the $\gamma$ corresponding to observational redshifts. Finally we 
find the corresponding $\chi_0^2$ 
values by using the formula 
\begin{equation}
\chi_0^2\,=\,\sum_{i=1}^{i=580}\{\frac{(\left(
\mu^{th(0)}(\gamma(i),r,s,\beta,\xi\xi_1,t_0)
-\mu^{obs}_i\right)^2}{\left(\sigma_i\right)^2}\} \label{mu3}
\end{equation}
where the subscript $0$ in $\chi_0$ and the superscript $(0)$ in 
$\mu^{th(0)}$ stands for the fact that $a(t)$ is 
approximated by its zero'th order approximation i.e. by $a_{1n}$, the 
superindices $th$ and $obs$ stand for the theoretical and 
observational values of $\mu$, and the subindices $i$ denote the values 
of the corresponding quantity for the $i$'th data point in 
Union2 data set. 

One may 
try a better approximation by replacing $a_{1n}(t)$ in (\ref{mu1b}) by a 
better approximation of
$a(t)$ i.e. by 
$\frac{c_0A_1}{1+c_0x^{(0)}(t)}a_{1n}(t)$ where $x^{(0)}(t)$ is defined by 
Eq.(\ref{a17a}).
In principle, then, one may evaluate the integral (\ref{mu1a}) after 
replacing $a_{1n}(t)$ by 
$\frac{c_0A_1}{1+c_0x^{(0)}(t)}a_{1n}(t)$.
However this seems to be 
inapplicable for 
standard computers because of the complicated form of the integral. One 
needs a separate computational physics project for this aim. Instead one 
may try a rough approximation (hopefully better than $a_{1n}$); we take 
the $\frac{1+c_0x^{(0)}}{c_0A_1}$ term in the integral to outside of the 
integral  
with its $\gamma$ value being the bound of the integral. This 
approximation is a good approximation provided that 
$\frac{c_0A_1}{1+c_0x(t)}$ does not vary much in the time interval between 
$t_0$ and the time corresponding to the given redshift value. Otherwise 
the higher order approximation may worsen the approximation rather than 
improving. 
The corresponding formulas (in the first order approximation) become
\begin{equation}
d_L\,\simeq\,
\left(\frac{1+c_0x^{(0)}}{c_0A_1}\right)^2\frac{c\,t_0\,
\beta^{-1+\gamma^{-\frac{1}{s}}}}{[1+\frac{\gamma-1}{r}\left(\xi\xi_1
-\frac{1}{s}\ln{\beta}\right)]^r}
\int_\gamma^1 
d\gamma\frac{\beta^{-1+\gamma^{-\frac{1}{s}}}}{[1+\frac{\gamma-1}{r}\left(\xi_1
-\frac{1}{s}\ln{\beta}\right)]^r}
\label{mu4}
\end{equation}
\begin{equation}
\chi^2\,=\,\sum_{i=1}^{i=580}\{\frac{(\left(
\mu^{th}(\gamma(i),r,s,\beta,\xi\xi_1,t_0)
-\mu^{obs}_i\right)^2}{\left(\sigma_i\right)^2}\} \label{mu5}
\end{equation}
After trial and error I have found many sets of parameters with relatively 
small $\chi_0^2$, $\chi^2$ values. For example the $\chi_0^2$,
$\chi^2$ values for two phenemonologically viable sets of parameters are 
given in Table \ref{table-9} where the reduced $\chi_0^2$,
$\chi_{red\,0}^2=\frac{\chi_0^2}{580-5}$, and the reduced $\chi^2$ values 
$\chi_{red}^2=\frac{\chi}{580-12}$ are in the order of 1 (where 580 is 
the number of data points, and 5, 12 are the number of free parameters $r$, 
$s$, $\beta$ etc. to be adjusted).

The sets of parameters (that I could by trial and error) with relatively small
$\chi^2$ values satisfy 
$c_1\simeq\,c_2\,\simeq\,1$, $\alpha_c\ll\,1$. By using this information
one may check the validity of (\ref{a10ba}) and determine if 
one may identify $\tilde{\Omega}_1$, 
$\tilde{\Omega}$, $\tilde{\Omega}_K$ by the corresponding density 
parameters; $\Omega_1$, $\Omega_x$, $\Omega_K$
for the phenomenologically relevant parameters 
by using Eq.(\ref{a10b}) and Eq.(\ref{a3db}) and 
Eq.(\ref{a3dc}). We observe that 
$x(0)$=$A_2$, $x_1(0)$=$x_2(0)$=$x_3(0)$=1, $c_1=1$ and 
for relevant values of the parameters 
Hence, after using (\ref{a10b}), we obtain
\begin{equation}
\frac{\tilde{\Omega}_x^\frac{1}{2}}{\Omega_b^\frac{1}{2}}\,\simeq\,
\left(\frac{\alpha_{o1}}{\alpha_{o1}+\alpha_{o2}}\right)
\left(\frac{\alpha_x}{\alpha_b}\right)~,~~~
\frac{\tilde{\Omega}_K^\frac{1}{2}}{\Omega_b^\frac{1}{2}}\,\simeq\,
\left(\frac{\alpha_{o2}}{\alpha_{o1}+\alpha_{o2}}\right)
\left(\frac{\alpha_K}{\alpha_b}\right)\label{mu7}
\end{equation}
We observe that for phenemonologically viable sets of parameters, for example, 
for those in 
Table \ref{table-9} we have 
$\tilde{\Omega}_x^\frac{1}{2}\,\sim\,\tilde{\Omega}_K^\frac{1}{2}\,\sim\,
\frac{1}{2}\Omega_b^\frac{1}{2}$ and
(\ref{a10ba}) may be satisfied since
$\tilde{\Omega}_1^\frac{1}{2}$=
$\xi_1\,\simeq\,0.98$$\sim\,1$. 
We notice that
$(\tilde{\Omega}_1^\frac{1}{2}
+\Omega_b^\frac{1}{2}+\Omega_r^\frac{1}{2}
+\tilde{\Omega}_x^\frac{1}{2}
+\tilde{\Omega}_K^\frac{1}{2})^2
\,\neq\,1$. However one may 
define a total density parameter for the dark sector by
\begin{equation}
\Omega_D^\frac{1}{2}=
\tilde{\Omega}_1^\frac{1}{2}
-\tilde{\Omega}_x^\frac{1}{2}
-\tilde{\Omega}_K^\frac{1}{2}
\end{equation}
Then the density parameters satisfies the necessary condition, 
$(\Omega_D^\frac{1}{2}+\Omega_b^\frac{1}{2}+\Omega_r^\frac{1}{2})^2=1$. 
In other words, $H_{1n}$ and $H_\Delta$ 
terms can not be identified as separate contributions to dark sector, 
rather they must be considered as just a single 
object in order not to introduce an ambiguity in their identification.

\subsection{Compatibility with Recombination and Nucleosynthesis}

In this subsection we investigate if this model is compatible with 
the 
cosmological depiction of the recombination and nucleosynthesis, at 
least, at the order of magnitude level. In a 
similar vein as the preceding subsection we require this model mimic the 
standard model, $\Lambda$CDM, as much as possible. We assume that the 
radiation 
and the baryonic matter are in thermal equilibrium in the eras of 
recombination and nucleosynthesis since we adopt the same equations of 
thermal equilibrium as $\Lambda$CDM. Therefore, in the following, first we 
drive the condition for thermal equilibrium for this model.  
Then we find the sets of parameters with least $\chi^2$ values that may 
produce 
successful recombination and nucleosynthesis eras. The correct choices 
should have sufficient radiation energy densities in these eras. In other 
words the redshift at the recombination time, $z_{re}$ should be 
in the order of $(1+z_{re})^4\,>\,
(1+z_*)^4\,\simeq\,(1100)^4$ where 
* denotes time of 
last scattering surface; and in the nucleosynthesis era
the energy density of neutrinos should reach energy densities of the order 
of $(1\,MeV)^4$. We seek an approximate, rough agreement with $\Lambda$CDM 
since the search of the parameter space is done by trial and error rather 
than a systematic search of the whole parameter space. Therefore a 
detailed, thorough analysis and compatibility survey would be too 
ambitious especially considering this is a toy model.

Before checking if there exist a set of parameters compatible with 
recombination and nucleosynthesis we should check if the thermal 
equilibrium is maintained in these eras in for the given set of parameters 
because we adopt the standard analysis in $\Lambda$CDM, and that analysis 
assumes existence of thermal equilibrium. As is well known, if there is 
thermal equilibrium then we should have $\Gamma\,>\,H$
where $\Gamma$ is the rate of the interaction between radiation and 
the matter and $H$ is the Hubble parameter. 
However the implementation of this condition in this model is not 
exactly the same 
as in $\Lambda$CDM.
In the case of recombination era the 
implementation of this condition does not give exactly the same result as 
$\Lambda$CDM since , in $\Lambda$CDM the recombination takes place in 
radiation dominated era and the total energy density is almost wholly 
due to radiation 
while, in this model, the total energy density of the universe at 
this era is not almost wholly due to radiation although the equation 
of state parameter for phenomenologically relevant cases is 
similar that of radiation dominated universe at the time of 
recombination and
we require the radiation 
energy density to be the same or almost the same as $\Lambda$CDM. In 
the case of nucleosynthesis, even the equation state parameter in 
this model does not mimic that of a radiation dominated universe. 
Therefore we should derive the corresponding conditions for thermal 
equilibrium for this model.

The condition for thermal equilibrium in the recombination era 
is
\begin{eqnarray}
&&\Gamma\,>\,H~\Rightarrow\,~1.97\times\,10^{-19}s^{-1}\times\,0.0227\,
(\frac{T}{T_{ph\,0}})\,>\,\alpha_1\,H_0\Omega_{ph}^\frac{1}{2} 
\nonumber \\
&&\Rightarrow~
\left(\frac{\frac{\rho}{\rho_0}}{\frac{\rho_{ph}}{\rho_{ph0}}}
\right)^\frac{1}{2}\,<\,
0.19\,\left(\frac{T}{T_{ph\,0}}\right) =0.19(1+z_{re})\nonumber \\
\label{rn5}
\end{eqnarray}
Here we have used the identities, 
\begin{equation}
\alpha_1^2=\frac{\rho}{\rho_{ph}}
=\frac{1}{\Omega_{ph}}\left(\frac{\frac{\rho}{\rho_0}}
{\frac{\rho_{ph}}{\rho_{ph0}}}\right) ~,~~
\frac{\rho_r}{\rho_{r0}}=\frac{\rho_{ph}}{\rho_{ph0}}=a^{-4} 
\label{rn5a} 
\end{equation}
where $\alpha_1\simeq\,1$ is the $\Lambda$CDM value and $\alpha_1^2\leq\,1$ 
at the time of recombination and is not constant in this model,
and I have used the PDG values, 
$H_0=72\,km\,Mpc^{-1}\,s^{-1}$, 
$\Omega_{ph}=4.8\times\,10^{-5}$. Note that $\Gamma$ in (\ref{rn5}) is the same 
as the $\Lambda$CDM value while $H$ is different from 
the $\Lambda$CDM value. 

Next consider the condition on thermal equilibrium at and before the time 
of nucleosynthesis. 
In thermal equilibrium we have
\begin{eqnarray}
&&\frac{\Gamma_\nu}{H}\,\approx\,\frac{1}{\alpha_2}\frac{\sqrt{45}G_{wk}^2}
{64\pi^3}\sqrt{\frac{hc^5}{G}}
(k_BT)^3
\,\simeq\,\frac{1}{\alpha_2}\left(\frac{T}{10^{10}K}\right)^3\,>\,1 
\nonumber \\
&&\Rightarrow\,
~~
\left(\frac{\frac{\rho}{\rho_0}}{\frac{\rho_r}{\rho_{r0}}}
\right)^\frac{1}{2}\,<\,
\Omega_r^\frac{1}{2}\left(\frac{T}{10^{10}K}\right)^3
\label{rn7}
\end{eqnarray}
where we have used the identity similar to (\ref{rn5a}), where $\alpha_1$ 
and the subindex $ph$ are replaced by $\alpha_2$ and $r$, respectively and the 
ratio is evaluated at the time of nucleosynthesis. In this case, as well, 
$\Gamma_\nu$ is the same as its $\Lambda$CDM value while the expression for 
$H$ in terms of temperature is different since $\alpha_2\leq\,1$ and is 
not a constant (i.e. it gives a different value when evaluated at different 
time during nucleosynthesis) in this 
model while $\alpha_2=1$ in $\Lambda$CDM. 
During 
thermal equilibrium the ratio of neutrinos to all nucleons, $X_n$ is given 
by
\begin{equation}
X_n\,=\,\frac{1}{1+\exp{(\frac{Q}{k_BT})}} \label{rn8})
\end{equation}
where $Q$ is the rest mass energy difference between a neutron and 
a proton, $Q=m_n-m_p=1.239\, MeV$. After the thermal equilibrium between 
the neutrinos and the nucleons are lost i.e. after decoupling the value of 
$X_n$ further decreases due to decay of free neutrons as
\begin{equation}
X_n\,=\,X_{n0}\exp{[-(\frac{t}{\tau_0})]} \label{rn9}
\end{equation}
where $X_{n0}$ is the $X_n$ of Eq.(\ref{rn8}) at the time of decoupling, 
and $\tau_0=885.7$ seconds is the lifetime of a free neutron. 
Therefore the effect of this model is to change the value of 
$X_{n0}$ (that depends on $\alpha_2$) and probably the 
value of $X_n$ as well.

Now we are ready to check the viability of this model.
I could give only four graphs 
and three tables that partially 
summarize the results of my calculations 
related to this and the next paragraphs in order 
not to expand the size of the paper too much. 
Otherwise the size of the manuscript would be 
almost doubled. First we check the viability of the 
model for recombination and nucleosynthesis eras. 
To this end I have used the equations 
(\ref{a11a},\ref{a16c},\ref{a3db},\ref{a3dc}) in the 
zeroth order approximation where 
$a(t)\simeq\,a_{1n}(t)$ (as discussed before Eq.(\ref{a3ba}))
to draw $\omega$, $\frac{\rho_r}{\rho_{r0}}$, 
versus time graphs by using a Mathematica code 
that I have prepared for this aim for the sets of the parameters, 
$r$,$s$,$\beta$,$\xi\xi_1$, $A_1$, $A_2$, $c_1$, $c_2$, 
$\frac{\alpha_r}{\alpha_b}$,$\frac{\alpha_x}{\alpha_b}$,
$\frac{\alpha_K}{\alpha_b}$, $\alpha_c$,$\alpha_{o1}$,
$\alpha_{o2}$, $t_0$ $\Omega_b^\frac{1}{2}$, that correspond 
to some relatively small $\chi^2$ 
values obtained in preceding subsection. 
Then I have tried to find at least one set of parameters with 
phenomenologically viable $\omega_0$, $\frac{\rho_r}{\rho_{r0}}$, 
$\frac{\rho}{\rho_0}$ values i.e. $\omega_0$, in the range 
$-0.68~$ \textemdash $~-0.74$; 
$\frac{\rho_r}{\rho_{r0}}\,>\,(1100)^2\simeq\,10^{12}$ (in the range 
of redshifts $z\sim\,800~-~3000$) at the time of
recombination, and
$\frac{\rho_r}{\rho_{r0}}\,>\,\frac{(1\,MeV)^4}
{5\times\,10^{-5}\,(2.5\times\,10^{-3}\,eV)^4}\,>\,10^{38}$
at the time of nucleosynthesis where I 
have approximated $x_1(t)$, $x_2(t)$, $x_3(t)$ 
by  $x_1^{(0)}(t)$, $x_2^{(0)}(t)$, $x_3^{(0)}(t)$ 
(that are defined in Eq.(\ref{a3ba})), and $a(t)$ 
by $a_{1n}(t)$ (that is defined in (\ref{a13b})) as 
discussed in the preceding section. I have found two sets 
of parameters given in Table \ref{table-9} that 
satisfy these conditions. 
A comment is in order at this point.
The zeroth order 
approximation is reliable only for small redshifts. However 
this approximation is reliable at any redshift if one is 
only interested in the energy density - redshift relation. 
This may be seen as follows: 
Assume that the energy density $\rho$ is related to redshift 
$z$ by $\rho\,=\,f(z)$ in the zeroth order approximation 
(where $f(z)$ is an arbitrary function), and
in an approximation 
better than the zeroth order we have
$\frac{c_0A_1}{A_1-A_2}=\frac{1}{x}$ i.e. $a(t)=\frac{1}{x}a_{1n}(t)$. 
Then the energy density after the correction 
is $\rho^\prime=f(z^{\prime})$. 
If one rescales $z^\prime$ as $\frac{1}{x}z^\prime=z$ then one 
obtains the same redshift and energy density values. In other words the 
redshift - energy density relation is invariant under such corrections. 
However this is not true for the redshift - time relation. If 
the approximation is not a good approximation to the true value then 
the redshift - time relation will be distorted. This, in turn, may cause 
the distortion of the value of the equation of state and the distortion 
of the variation of the energy densities with time in an amount 
depending on the reliability of zeroth order approximation. Keeping these 
observations in mind I content to use zeroth order approximation for the times 
of recombination and nucleosynthesis because even employing zeroth order 
approximation needs a lot of computer CPU and RAM, and in many cases 
the use of first order approximation as well does not improve the situation. 
We will come back to these points when discuss the times of recombination and 
nucleosynthesis.

\begin{figure}
\includegraphics[scale=0.9]{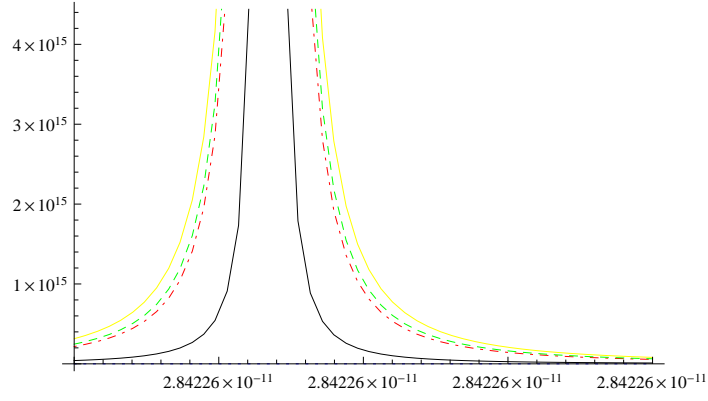}
\caption{$\omega$ (dotted blue), 
$\frac{\rho}{\rho_0}$ (dot-dashed red), 
$\frac{\rho_x}{\rho_{x0}}$ (solid black), 
$\frac{\rho_r}{\rho_{r0}}$ (dashed green), 
$\frac{\rho_b}{\rho_{b0}}$ (solid yellow) 
versus $\gamma=\frac{t}{t_0}$ 
graphs for the set B in Table 9 for the first energy density peak 
in the interval 
$2.8422577892
\times\,10^{-11}\,\leq\,\gamma\leq\,
2.8422577894\times\,10^{-11}$.
In this graph $\omega$, $\frac{\rho}{\rho_0}$,
$\frac{\rho_r}{\rho_{r0}}$, $\frac{\rho_b}{\rho_{b0}}$ are given as 
multiples of $10^4$, 
$10^{28}$, $10^{31}$, $10^{23}$, $10^{19}$, respectively.}
\label{fig:9}
\end{figure}

\begin{figure}
\includegraphics[scale=0.9]{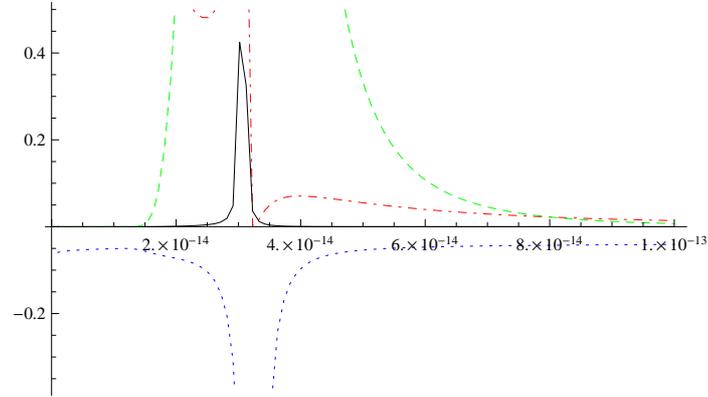}
\caption{$\omega$ (dotted blue), 
$\frac{\rho}{\rho_0}$ (dot-dashed red), 
$\frac{\rho_x}{\rho_{x0}}$ (solid black), 
$\frac{\rho_r}{\rho_{r0}}$ (dashed green), 
$\frac{\rho_b}{\rho_{b0}}$ (solid yellow) 
versus $\gamma=\frac{t}{t_0}$ 
graphs for the set B in Table 9 for the second energy density peak 
in the 
interval 
$10^{-15}\,\leq\,\gamma\leq\,
10^{-13}$.
In this graph 
$\omega$, 
$\frac{\rho}{\rho_0}$,
$\frac{\rho_r}{\rho_{r0}}$,
$\frac{\rho_b}{\rho_{b0}}$ are given as multiples of 10, $10^{28}$, 
$10^{31}$, $10^{27}$, $10^{20}$, respectively. }
\label{fig:10}
\end{figure}

\begin{figure}
\includegraphics[scale=0.9]{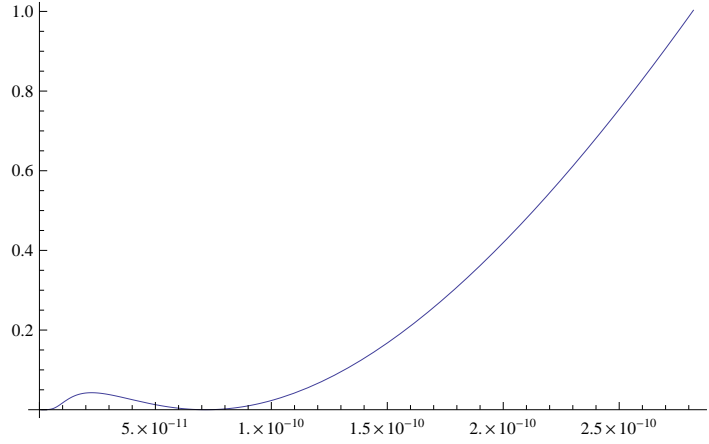}
\caption{
$\left(
\frac{\frac{\rho}{\rho_0}}{\frac{\rho_r}{\rho_{r0}}}\right)^\frac{1}{2}
\left(\frac{1}{0.19
\frac{T}{T_{ph0}}}\right)$ versus $\gamma$ graph in the interval where 
thermal equilibrium is maintained
for the set B in Table \ref{table-9}
in the 
interval 
$10^{-12}\,
\leq\,\gamma\leq\,
2.82\times\,10^{-10}$ }
\label{fig:15}
\end{figure}

Next I have checked if thermal equilibrium is maintained at the times of 
recombination and nucleosynthesis and if recombination and nucleosynthesis are
realized in this model. 
One may have an idea on thermal equilibrium at the time of recombination by 
using the values of Table \ref{table-9} at $z\simeq\,1100$ and 
Eq.(\ref{rn5}). However a more rigorous way is to draw
$\left(
\frac{\frac{\rho}{\rho_0}}{\frac{\rho_r}{\rho_{r0}}}\right)^\frac{1}{2}
\left(\frac{1}{0.19
\frac{T}{T_{ph0}}}\right)$ (that may be obtained from 
Eq.(\ref{rn5})) versus time graphs to determine the time intervals (and 
then the corresponding redshift intervals) where 
$\left(\frac{\frac{\rho}{\rho_0}}{\frac{\rho_r}{\rho_{r0}}}\right)^\frac{1}{2}
\left(\frac{1}{0.19\frac{T}{T_{ph0}}}\right)\leq\,1$ for each of the sets 
A and B. In fact I have used $\frac{T}{T_{ph0}}\,=\,1+z$ for the 
relevant redshifts. The resulting intervals are the intervals where thermal 
equilibrium is maintained as shown in Figure \ref{fig:15} for the set B 
in Table \ref{table-9}. The smallest redshifts where the thermal 
equilibrium is lost are $z=2317$ ($\gamma=1.615\times\,10^{-10}$ 
with $\frac{\rho_r}{\rho_{r0}}\,\simeq\,5.52\times\,10^{13}$) and  
$z=1625$ ($\gamma=2.82\times\,10^{-10}$
$\frac{\rho_r}{\rho_{r0}}\,\simeq\,3\times\,10^{13}$) for the sets A and B, 
respectively. This implies that the photon electron decoupling 
takes place before the time of last scattering at an energy of 
$\simeq\,2317\times\,6\times\,10^{-4}$ eV$\,\simeq\,1.4$ eV and   
$\simeq\,1625\times\,6\times\,10^{-4}$ eV$\,\simeq\,0.98$ eV for the sets 
A and B, respectively (assuming the transition being directly to the 
ground state of hydrogen atom) to be compared to the value of photon energy 
of about $\simeq\,1100\times6\times\,10^{-4}$ eV$\,\simeq\,0.66$ eV for 
$\Lambda$CDM at the time of last scattering.
This, in turn, implies that photon electron decoupling 
in this model for the sets of 
parameters A and B is at a smaller redshift than 
$\Lambda$CDM where thermal equilibrium is maintained until decoupling. 
(Thermal equilibrium would be maintained till  $z\simeq\,2.4$ in 
$\Lambda$CDM if recombination of electrons and protons to form neutral 
atoms had not taken place as may be seen from (\ref{rn5}) by setting 
$\alpha_1=1$). In fact the corresponding times for decoupling are already 
smaller than that of $\Lambda$CDM by five orders of magnitude. A detailed 
comprehensive separate study is need to see if these imply some 
interesting phenomenologically viable alternatives or just an artifact of the toy 
model and/or the sets of parameters considered. This may also be due to the limitation 
of the applicability of zeroth order approximation that we have discussed above.
$a(t)\simeq\,a_{1n}(t)$ is not violated badly at the 
time of recombination for the most of the relevant sets of
parameters. For example for the sets of parameters given in
Table \ref{table-9} the first order approximation results in
$a(t)\simeq\,0.4\,a_{1n}(t)$ i.e. 
$\frac{c_0A_1}{A_1-A_2}\simeq\,0.4$ and does not vary much at  
the time of recombination. Therefore it seems that the effect of the 
limitation of the applicability of zeroth order approximation to the time of recombination 
must be limited. 
However this shift does not 
introduce a major problem since the redshift values, hence 
the photon energy density at recombination remains almost the same 
and thermal equilibrium is maintained.   
Next I have checked if thermal equilibrium is maintained
at the peaks in Table \ref{table-9} where the energy densities 
are sufficient for nucleosynthesis. 
I have used Eq.(\ref{rn7}) to find the range of 
temperatures where thermal equilibrium is maintained.
I have found that this condition is 
satisfied for $T\,>\,3\times\,10^{10}\,K$ (provided that 
$\Omega_r\simeq\,5\times\,10^{-5}$) for the second peaks. This value gives 
us $X_{n0}$ in (\ref{rn9}) by using (\ref{rn8}) as $X_{n0}\simeq\,0.39$ 
which is quite large compared to the $\Lambda$CDM value of 
$\simeq\,0.25$. The time that takes $3\times\,10^{10}K\simeq\,1\,MeV$ 
drop to 0.07 MeV (that is when     
$\frac{\rho_r}{\rho_{r0}}\,\sim\,10^{32}$) in this model is something like 
$\sim\,2\times\,10^{-16}\times\,t_0\simeq\,90$ seconds. Therefore 
$X_{n0}$ does not drop significantly through Eq.(\ref{rn9}). In other 
words the final result $X_{n}\simeq\,0.35$ is much larger than the
$\Lambda$CDM value $\simeq\,0.13$ (which agrees well with observations). 
Probably the main source of this discrepancy is inapplicability of zeroth order
approximations to redhifts and energy densities to this era to obtain 
correct  energy density - time relations. 
The variations of $\frac{c_0A_1}{A_1-A_2}$ and $A$ are quite 
large and their values are quite different
than those at $z\sim\,0$ 
at the time of nucleosynthesis that makes the 
applicability of the zeroth order approximation extremely difficult to 
obtain correct energy density - time relation. In other words the 
main source of the discrepancy may be due to the fact that 
the real time of free decay 
may be of the order of $\sim\,1000$ seconds in this model 
instead of 90 seconds.
The use of first order approximation does not improve the situation
because in the calculation of first order approximations to
$\frac{c_0A_1}{A_1-A_2}$ and $A(t)$
one uses the zeroth order approximation $a(t)\,\simeq\,a_{1n}(t)$ in 
the integrals for $x_i$, $i=1,2,3$. This, in turn, results in 
over contribution of large redshifts and hence 
larger and more varying $x_i$ with respect to their true values since
$x_i\,<1$ and they get smaller i.e. 
$\frac{c_0A_1}{A_1-A_2}$ gets larger
at larger redshifts. 
Therefore the energy density versus time graphs in
the figures \ref{fig:9}, \ref{fig:10} and \ref{fig:15} must 
be considered with some care: The time values in those graphs
should be taken with utmost care 
especially in the case of nucleosynthesis
while the magnitudes of energy 
densities and the corresponding redshifts are expected to 
be the same as the exact values.
All these points must be studied in more detail in future studies.
However I have been able to show that this scheme can produce a model
that mimics the standard model: 
There is a current accelerated epoch whose present equation of 
state (for the whole universe) is ~-0.7 (that is, at least, 
roughly in agreement with observations e.g. see the value in Table \ref{table-9} 
for a phenomenologically relevant set of parameters). 
Before this epoch $\omega$ changes sign 
and the time near this sign change may be considered as the matter 
dominated era. Although the sign change of $\omega$ occurs at a later time 
in this model compared to $\Lambda$CDM the time and the 
redshift of onset of the 
accelerated era (i.e. $\omega\simeq\,-\frac{1}{3}$) are comparable with 
those of $\Lambda$CDM.
There is an epoch before the matter dominated era 
where $\omega$ is on average close to 
$\frac{1}{3}$, and may be identified by radiation dominated era, 
and the time of the maximum value of $\omega$ may be considered as the time when 
the universe was like stiff matter or denser (as 
in the cores of stars). Then $\omega$ changes sign 
again reaches to two minima peaks as mentioned before and eventually 
approaches to -1 as time goes to zero 
(due to the $H_{1n}$, in particular the first part of 
it) and 
this epoch probably may be considered as the inflationary era. 
Moreover the model is able to give relatively small reduced 
$\chi_0^2$ and $\chi^2$ values for Union2.1 data set, and it 
can, at least roughly, account for 
recombination and nucleosynthesis times. I think 
this is a sufficiently well starting point for a toy model whose 
main aim is to embody the creation of matter 
and radiation in the scale 
factor of Robertson-Walker metric. However 
there is a great deal of points to be clarified and addressed in future 
studies such as checking the whole parameter space of this model by using 
a more elaborate software and to use more powerful computers that may 
give scan the whole parameter space in a better approximation than the one 
given here, and considering a more detailed analysis of the recombination 
and nucleosynthesis epochs, studying the evolution of cosmological 
perturbations in this model, and considering 
possible extensions of this model towards a more realistic model. 

\section{Conclusion}

In this study a scheme for obtaining a scale factor (in Robertson-Walker 
metric) that may account for the times before, during, and after the 
radiation dominated eras is introduced. The prescription to 
obtain the scale factor in this model is quite simple; First one  
introduces a scale factor for the pure dark sector, and then the full 
scale factor is obtained by a relation between these two scale factors. 
The 
result is a scheme to produce the scale factor for the whole universe, 
including baryonic matter, radiation, and dark energy-matter (i.e. dark 
sector) in such a way that the times before, during, and after radiation 
dominated era are expressed by a single scale factor in Robertson-Walker 
metric. Different choices of the pure dark sector scale factor (denoted 
by $a_1$ in this paper) and different choices of the relation between 
$a_1(t)$ and the scale factor of the full universe, $a(t)$ give different 
models. As an illustration of this scheme a model with a specific scale 
factor for the 
pure dark sector and a specific relation between $a_1(t)$ and $a(t)$ is 
considered. The phenomenological viability of this model is checked 
through its 
compatibility with Union2.1 data set, and with recombination and 
nucleosynthesis by using trial error and Mathematica 
software for almost randomly chosen sets of parameters. Two sets of 
parameters with relatively small $\chi^2$ values for Union2.1 data set, 
and that are compatible with successful recombination and nucleosynthesis 
at an order of 
magnitude level are found. These results are encouraging in view of the 
fact that only a tiny portion of the whole parameter space could be 
considered in this way. A separate, detailed, and comprehensive 
computational 
project with more advanced software codes and/or powerful 
computing facilities that may scan the full parameter space and may employ 
better approximation schemes is needed to reach a definite view on the 
observational viability of this scheme and/or this model. Moreover the 
effect of this model on cosmological perturbations should be considered 
and possible implications and extensions of this 
scheme to inflationary era should be studied in future. Furthermore 
different pure dark sector scale factors and different options to relate 
pure dark sector  and the full universe scale factors may be 
considered in future to see the full range of possibilities that this 
scheme may offer.  

\begin{acknowledgments}
I would like to thank Professor Joan Sol$\grave{a}$ for reading the manuscript 
and for his valuable comments. I would also like to thank Professor 
Gregory Gabadadze for reading the manuscript, and to
Dr. A. Aviles for his help in the initial 
phase of writing the Mathematica code used in this study.
\end{acknowledgments}

\appendix

\section{Mathematica Codes to Evaluate $\chi^2$ and to Draw the Plots}

In this appendix I give the essential, non-trivial steps for
writing the Mathematica codes to find $\chi_0^2$, $\chi^2$, and 
to plot the graphs for the equation of state and the energy densities 
in this model. To find $\chi_0^2$ or $\chi^2$ 
in Subsection IV.A we first find the distance moduli given by 
\begin{equation}
\mu\,=\,5\,Log_{10}\left(\frac{d_L}{1\,Mpc}\right)\,+\,25 
\end{equation}
Here
\begin{equation}
d_L\,=\,\frac{c\,a_0}{a(t)}\int_t^{t_0}\frac{dt}{a(t)}\,\simeq\,
\frac{c\,a_0}{
a_{1n}(t)\zeta^{-1}}
\int_t^{t_0}\frac{dt}{\zeta^{-1}a_{1n}(t)} 
\label{apd1}
\end{equation}
where $\zeta^{-1}=1$ in the calculation of $\chi_0^2$ and
$\zeta^{-1}=\frac{A_1c_0}{1+c_0x^{(0)}}$ 
in the calculation of $\chi^2$ of Subsection IV.A. 
In general $d_L$ is written in terms of Hubble parameter $H$ and redshifts 
since it is more suitable for
the analysis of the data, which are given as distance moduli 
at various redshifts. 
On the other hand, in this case, we do not express $d_L$ in terms of redshift
because expressing Hubble parameter as a function of redshift is not 
applicable in this case due to the complicated form of the scale factor. 
Instead we convert the redshift values in Union2.1 data set into 
time by setting \\
$\gamma$a$\gamma$inv[r$\_$,s$\_$,b$\_$,z1$\_$,z$\_$]:= \\	
a$\gamma$/.Table[FindRoot[a$\gamma$inv[a$\gamma$,r,s,b,z1,z]
-Union2z$\mu$error[[i,1]]-1
==0,{a$\gamma$,0.1}],{i,1,numberUnion2}];\\ 
where 
$\gamma$a$\gamma$inv[r$\_$,s$\_$,b$\_$,z1$\_$,z$\_$],  
a$\gamma$inv[a$\gamma$,r,s,b,z1,z], Union2z$\mu$error[[i,1]] stand for
$\gamma=\frac{t}{t_0}$, $\frac{1}{a}\simeq\,\frac{1}{\zeta^{-1}a_{1n}}$, and the redshift
for the i'th data in Union2.1 data set; respectively. Then $\chi_0^2$ or $\chi^2$ is 
calculated by numerical integration by Mathematica through the formula\\
$\chi$[r$\_$, r1$\_$, s$\_$, s1$\_$, $\beta\_$, b$\_$, $\xi1\_$,
z1$\_$, z$\_$, Ho$\_$]:=
Sum[(($\mu$t[$\gamma$a$\gamma$inv[r, 
s, b, z1, z][[i]], r1, s1, $\beta$, $\xi$1, z, Ho]
-$\mu$U2[i])$\wedge$2)/($\sigma$U2[i])$\wedge$2,{i,1,numberUnion2}]; \\
where we take $a(t)=a_{1n}(t)$ for $\chi_0^2$ and 
$a(t)=\frac{A_1c_0}{1+c_0x^{(0)}}a_{1n}(t)$ for $\chi^2$, $\mu$U2[i] 
is the magnitude for the i'th data, $\sigma$U2[i] is the error for 
the i'th data in Union2.1. 
$\frac{1}{a_{1n}}$, for example may be expressed
as\\
inv[r$\_$, s$\_$, b$\_$, z1$\_$, z$\_$, g$\_$] := 
(1 +(1/r)*(g - 1)*(z1*z - (1/s)*Log[b]))$\wedge$(-r)
*b$\wedge$(g$\wedge$(-1/s) - 1).
In the case of $\chi^2$ one should also 
write the expressions for $x_i^{(0)}$ to find $x^{(0)}$ before 
evaluation of $\chi^2$. 
To draw the graphs we write expressions for the Hubble parameters
due to each contribution. This may be done for $H_{1n}$ by using 
$H_{1n}=\frac{\dot{a}_{1n}}{a_{1n}}$.
For the other components, for example for
the dust component, by
evaluating $A^{(0)}\frac{\alpha_b}{a_{1n}
^\frac{3}{2}}$. We use the fact 
$\frac{\rho}{\rho_0}=\frac{H^2}{H_0^2}$ to draw the related graphs.

\bibliographystyle{plain}

\newpage

\begin{table}[h] 
\begin{tabular}{|c|c|c||c|c|c||c|c|c||c|c|c||c|c|c|} 
\hline
$\gamma$&$A^{(0)}$&$A^{(1)}$&$\gamma$&$A^{(0)}$&$A^{(1)}$&
$\gamma$&$A^{(0)}$&$A^{(1)}$&$\gamma$&$A^{(0)}$&$A^{(1)}$&
$\gamma$&$A^{(0)}$&$A^{(1)}$
\\
\hline 
2&2.9352&2.9679&1.7&2.9453&2.9679
&1.4&2.9494&2.9679&1.2&2.9468&2.9679&
1&2.9383&2.9635 \\
\hline 
0.9&2.9317&2.9635&0.8&2.9235&2.9635
&0.7&2.9142&2.9635&0.6&2.9050&2.9635
&0.5&2.8983&2.9635 \\
\hline 
0.4&2.8988&2.9635&0.3&2.9169&2.9635&
0.2&2.9766&2.9635&0.1&3.1490&2.9635&
$10^{-8}$&4.5020&-
\\
\hline 
$10^{-10}$&5.8496&-&$10^{-11}$&-4.6576&-
&$10^{-12}$&-2.2796&-&$10^{-14}$&$\sim\,10^{-7}$&-&
$10^{-16}$&$\sim\,0$&-
\\
\hline\hline\hline 
\end{tabular}
\caption[b]{The zeroth and the first 
order approximations to $A(t)$;
$A^{(0)}(t)$ and $A^{(1)}(t)$
as function $\gamma=\frac{t}{t_0}$ 
for the set of parameters, 
$r$=.138, $s$=12, $\beta$=3.3, 
$\xi_1\xi$=0.975, $A_1$=1, $A_2$=0.002745, $c_1$=1, $c_2$=0.9986,
$\frac{\alpha_r}{\alpha_b}$=0.03, $\alpha_c$=$10^{-7}$, 
$\alpha_{o1}$=0.5, $\alpha_{o2}$=1,
$\frac{\alpha_x}{\alpha_b}$=1,
$\frac{\alpha_K}{\alpha_b}$=0.8, $\Omega_b^\frac{1}{2}=0.22$. Note that
first order value $A^{(1)}$ for $\gamma=10^{-8}$ and smaller values of 
$\gamma$ are not evaluated since
the iteration procedure is not applicable for such small
times because of the time intervals, 
$t_i-t_{i-1}=0.1\,t_0$ that we have used in Eq.(\ref{ax}) 
is much coarser than $10^{-8}$.} 
\label{table-1}
\end{table}

\begin{table}[h] 
\begin{tabular}{|c|c|c|c|c|} 
\hline
$\gamma$
&1+$10^{-8}$&0.9+$10^{-8}$&0.8+$10^{-8}$&0.7+$10^{-8}$
\\
\hline
$\zeta$&0.9999999996313257&1.0034967688351344&1.0062228502031836
&1.0072827970804177\\
\hline
$\gamma$
&0.6+$10^{-8}$
&0.5+$10^{-8}$
&0.4+$10^{-8}$
&0.3+$10^{-8}$
 \\
\hline
$\zeta$&1.0051401822527948&0.9971498327626435 
&0.9787034586841717&0.941589630495205 \\
\hline\hline\hline 
\end{tabular}
\caption[b]{$\zeta(t)$=$\frac{(1+c_0x)}{c_0A_1}$ versus $\gamma=\frac{t}{t_0}$ values 
for the set of  the parameters; $r$=2.138, $s$=12, $\beta$=3.3, 
$\xi_1\xi$=0.975, 
$A_1$=$c_1$=$\alpha_{o2}$=1, $A_2$=0.002745, $c_2$=0.9986, 
$\alpha_{o1}$=0.5,
$\frac{\alpha_r}{\alpha_b}$=0.03, $\alpha_c=10^{-7}$, 
$\frac{\alpha_x}{\alpha_b}$=1, $\frac{\alpha_K}{\alpha_b}$=0.8, 
$t_0=\,\frac{1}{72.8}$ (Mpc/km) s, $\chi_0^2$=579.97, 
$\chi^2$=576.69. Note that $\frac{\zeta_{av}}{\zeta_0}\simeq\,$0.9996 
is rather close to 1 where $\zeta_{av}$, $\zeta_0$ are the average value 
of $\zeta$ and the value of $\zeta$ at $\gamma=1$, respectively.}
\label{table-7} 
\end{table}

\begin{table}[h] 
\begin{tabular}
{|c|c|c|c|c|c|c|c|c|c|c|c|c|c|c|} 
\hline
Set&$\chi^2$&$\omega_0\simeq$
&$\gamma_{rc}\simeq$&$z_{rc}\sim$
&$\left(\frac{\rho_r}{\rho_{r0}}\right)_{rc}$
&$\left(\frac{\rho}{\rho_{0}}\right)_{rc}$&$\omega_{rc}$ \\
\hline
A&623.203&-0.69&$5\times\,10^{-10}$&1148 
&$2.9\times\,10^{12}$&$1.18\times\,10^{18}$&0.245 \\
\hline \hline
A&$\gamma_{ns}\simeq$&$z_{ns}\sim$
&$\left(\frac{\rho_r}{\rho_{r0}}\right)_{ns}$
&$\left(\frac{\rho}{\rho_{0}}\right)_{ns}$
&$\omega_{ns}$
&$\omega_{min}\simeq$&$\omega_{max}$\\
\hline 
A&$1.5\times\,10^{-11}$&
12839&$10^{37}$&$9.6\times\,10^{41}$
&-1.67&$-5\times\,10^{29}$($\gamma\simeq\,3.7\times\,10^{-11}$)
&6.22 \\
&$3.1\times\,10^{-14}$&$7.9\times\,10^6$&
$2.7\times\,10^{37}$
&$8.4\times\,10^{35}$&-8.5 & 
$-5\times\,10^{30}$ 
($\gamma\simeq\,3.3\times\,10^{-14}$)& 
($\gamma\simeq\,0.015$)\\
\hline \hline \hline \hline
Set&$\chi^2$&$\omega_0$
&$\gamma_{rc}\simeq$&$z_{rc}\sim$
&$\left(\frac{\rho_r}{\rho_{r0}}\right)_{rc}$
&$\left(\frac{\rho}{\rho_{0}}\right)_{rc}$
&$\omega_{rc}$ \\
\hline
B&576.69
&-0.6948&
$5\times\,10^{-10}$&1148
&$6.8\times\,10^{12}$
&$10^{18}$
&0.257 \\
\hline \hline
B
&$\gamma_{ns}\simeq$&$z_{ns}\sim$
&$\left(\frac{\rho_r}{\rho_{r0}}\right)_{ns}$
&$\left(\frac{\rho}{\rho_{0}}\right)_{ns}$
&$\omega_{ns}$
&$\omega_{min}\simeq$&$\omega_{max}$\\
\hline
B&$2.84\times\,10^{-11}$& 
7827&$1.3\times\,10^{45}$&$1.1\times\,10^{50}$&-1.67&  
$-1.5\times\,10^{29}$ 
($\gamma\simeq\,7.2\times\,10^{-11}$)
&$28$ \\
&$3.1\times\,10^{-14}$& 
$8\times\,10^6$
&$1.5\times\,10^{42}$ 
&$4.9\times\,10^{40}$
&-9.6&  
$-2.8\times\,10^{25}$ 
($\gamma\simeq\,2.3\times\,10^{-14})$&
($\gamma\simeq\,0.029$)\\
\hline
\end{tabular}
\caption[b]{Some of the sets of parameters with sufficient energy 
densities for recombination and nucleosynthesis with relevant redshift 
values. Here 
$\gamma_{rc}=\frac{t_{rc}}{t_0}$,
$\gamma_{ns}=\frac{t_{ns}}{t_0}$; the subscripts, $rc$ and $ns$ denote  
recombination and nucleosynthesis, respectively;
A$\equiv\,$($r=2.138$, $s=12$, $\beta=3.3$, $\xi\xi_1=0.975$, $\xi=1$, 
$\Omega_b^\frac{1}{2}=0.22$, $A_1=1$, $A_2=0.002745$, 
$c_1=1$, $c_2=0.9986$, $\frac{\alpha_r}{\alpha_b}=0.03$, 
$\alpha_c=10^{-7}$, $\alpha_{o1}=0.3$, $\alpha_{o2}= 
1$, $\frac{\alpha_x}{\alpha_b}=1$, $\frac{\alpha_K}{\alpha_b}=0.8$). The 
set B is the same as the set A except $\alpha_{o1}$ is replaced by 0.5.
The best $\chi^2$ values for the sets A 
and B correspond to 
$t_0=\frac{1}{72.9\,km\,s^{-1}\,Mpc^{-1}}\,\simeq\,13.25$ years and 
$t_0=\frac{1}{72.8\,km\,s^{-1}\,Mpc^{-1}}\simeq\,13.3$ years, 
respectively. 
Note that the shape of the $\omega$ versus $\gamma$ is 
extremely sharp time hence the 
location of $\left(\frac{\rho_{r}}{\rho_{r0}}\right)_{ns}$ is 
sensitive to exact value of $\gamma$. The more exact values of 
$\gamma$ for the set A and B where $\omega$ is minimum are  
($\gamma\simeq\,3.67027978907089\times\,10^{-11}$,
$\gamma\simeq\,3.31606438849865\times\,
10^{-14}$)
and ($\gamma\simeq\,7.1641402601557\times\,10^{-11}$,
$\gamma\simeq\,2.246997272913787\times\,10^{-14})$), respectively. A 
similar case is true for $\gamma_{ns}$ since it is extremely small.
The more exact values of $\gamma_{ns}$ for the set A and B are  
($1.498712948305\times\,10^{-11}$, $3.1102\times\,10^{-14}$), 
($2.8422577892672\times\,10^{-11}$, $3.06971839\times\,10^{-14}$), 
respectively.}
\label{table-9} \end{table}


\end{document}